\documentstyle[preprint,prl,aps]{revtex}

\begin{document}
\draft

\title{
Fractal Dimension of Disordered Submonolayers: Determination from He
Scattering Data
}

\author{D.A. Hamburger$^{a,b}$, A.T. Yinnon$^{b}$ and R.B. Gerber$^{b,c}$}
\address{
$^a$Department of Physics, The Hebrew University of Jerusalem, Jerusalem
91904, Israel\\
$^b$The Fritz Haber Center for Molecular Dynamics, The Hebrew University of
Jerusalem, Jerusalem 91904, Israel\\
$^c$Department of Chemistry, University of California - Irvine, Irvine, CA
92717, USA
}

\maketitle

\begin{abstract}
\newline{}
We propose a novel method to measure the fractal dimension of a submonolayer
metal adatom system grown under conditions of limited diffusivity on a
surface. The method is based on measuring the specular peak attenuation of He
atoms scattered from the surface, as a function of incidence energy. The
(Minkowski) fractal dimension thus obtained is that of contours of constant
electron density of the adatom system. Simulation results are presented, based
on experimental data. A coverage dependent fractal dimension is found from a
two-decade wide scaling regime.\\
\end{abstract}

\newpage

\section{Introduction}
Fractal dimension (FD) \cite{Mandelbrot} has proved to be a very useful
concept for the description of certain disordered systems
\cite{Vicsek-Bunde,Avnir:book}. The experimental determination of this
quantity is therefore an important challenge. Scanning tunneling microscope
(STM) studies of adsorption and homoepitaxial growth of metal atoms on metal
surfaces have recently revealed the formation of fractal-like structures,
under conditions of limited adatom diffusivity and low deposition flux
\cite{Comsa:STM1,Behm,Kern:2}. The characterization of the structures formed
during these growth processes is a major objective of surface science, and the
determination of FD plays an important part in meeting this objective.  At
present, STM is virtually the only available method for measuring the FD of
such 2D systems. This involves image analysis of the STM topograph, as
performed, e.g., by Hwang et al. \cite{Behm} for Au/Ru(0001) and Brune et al.
\cite{Kern:2} for Ag/Pt(111). The STM method has, however, several
experimental limitations. In particular, the method is inherently local, and
the statistics for characterizing disordered surfaces, including fractal
structures, may not be the best. In general, to probe these systems at the
microscopic level requires resolution of several \AA.  Scattering techniques
used conventionally for the measurement of fractal dimension at such high
resolutions, such as X-rays and neutron diffraction \cite{Teixeira}, are not
applicable, due to their penetration into the bulk. An attractive alternative
is presented by He scattering, which is particularly well suited to the study
of disordered surfaces \cite{Comsa:book} as it is not destructive, probes
solely the surface, and provides excellent statistics: A single beam can cover
an entire surface of close to a cm$^2$ at once, compared to merely several
$(10^3\AA)^2$ with STM.

In the present Letter, we propose a novel method (SIMF - Specular Intensity
Measurement of Fractality) by which a He scattering experiment can be used to
measure the FD of submonolayer disordered surfaces. In sharp contrast to STM,
which due to its inherent local nature observes {\em individual clusters}, the
proposed method {\em measures the FD of an entire disordered surface}. Thus
the fractal property observed by a SIMF experiment, differs significantly from
the STM one. Another difference concerns the physical interpretation of what
is actually measured: STM probes the electron density near the Fermi level. In
the repulsive regime, He experiences an exchange interaction with other
electrons: It is scattered from regions that represent typically the low
density tail of the surface electrons. In a semi-classical picture, at a given
incidence energy, He is scattered from contours of constant potential. Using
the Esbjerg-Norskov-Lang \cite{Norskov} relation

\begin{equation}
V_{rep}(x,y,z) = \alpha\,\rho(x,y,z)
\label{eq:Norskov}
\end{equation}

\noindent for the {\em repulsive} part of a He/metal interaction, where
$\alpha$ is a system-dependent constant and $\rho(x,y,z)$ the surface electron
density, it follows that {\em it is the FD of constant electron density
contours which is measured in a SIMF experiment}. These two differences
illustrate that SIMF, which offers a {\em global} view at the {\em electron
density tail level}, is in fact complementary to the measurement of FD by STM,
which measures locally the electron density at the Fermi level. More
explicitly, since SIMF looks at contours of constant and low electron density,
it observes an object which is quite different from the underlying geometrical
set of adatom positions. This object is the ``thickened set'', or Minkowski
Sausage'' \cite{Mandelbrot} of the adatoms (see inset of
Fig.\ref{fig:D}). SIMF is simple to implement experimentally: All it requires
is the measurement of specular intensities, over a wide range of incidence
energies. This has to be performed for calibration on a very low coverage
system (ideally a single adatom), and then on the system of interest itself. A
straightforward analysis subsequently yields the FD. The method is based on
the energy-dependence of the scattering cross section (CS) of He scattering by
an adatom \cite{Comsa:book,me:optical}.  This dependence can be exploited to
observe the adatom system at different resolutions, thus providing the
variation of yardstick sizes necessary for FD measurements.

\section{Theory}
Consider first the measurement of the CS of a {\em single} adatom adsorbed on
an otherwise flat surface (as is the case to a good approximation for
Pt(111)). Assume also that the surface is perfectly rigid. Any off-specular
intensity must be attributed to scattering by the adatom
\cite{Comsa:book}, which allows for a convenient measurement of
the CS as

\begin{equation}
I(k_z) = I_0 \,e^{-n\,\theta\,\Sigma_s(k_z)}.
\label{eq:I-single}
\end{equation}

\noindent Here $I$ and $I_0$ are, respectively, the specular intensities for
scattering from a surface with a single adatom and a flat surface; $n$ is the
number of sites per unit-cell; $\theta$ is the coverage; $\Sigma_s$ is the
adatom CS; and $k_z$ is the incoming He wavenumber, assumed for simplicity to
strike the surface at normal incidence. One can accordingly associate a {\em
scattering radius}, $R_s(k_z) = (\Sigma_s(k_z)/\pi)^{1/2}$ with a single
adatom. (Formally Bragg scattering must be counted along with specular
scattering, but this effect is negligible for the flat metal surfaces
considered here. We have shown elsewhere \cite{me:optical} that
Eq.(\ref{eq:I-single}) is rigorously equivalent to the formal definition of
the scattering cross section of a {\em single} adatom). Next consider the CS
measurement of a disordered submonolayer of adatoms. Suppose that every collision with
an adatom results in off-specular scattering. This may be the case if the
coverage is sufficiently low, so that no large compact regions have formed; or
if the ad-islands are ``dendritic'' (fractal), so that the width of their
``arms'' is only several atoms across; or if the adatoms are non-metallic
(e.g., CO), so that they present a highly corrugated surface to the incident
He beam. Then the probability of hitting
the adatom system is just the off-specular intensity. But this is also clearly
the fraction of surface occupied by the adatom system. Thus

\begin{equation}
A_s(k_z) = 1-I(k_z),
\label{eq:I-system}
\end{equation}

\noindent where $I$ is now the specular intensity for scattering from the
adatom {\em system}, and $A_s$ is the total CS of the diffusely scattering
adatoms, normalized to the total available surface area. Thus $A_s$ cannot be
considered the total CS of large {\em compact} metallic islands, which scatter
diffusely only from their rim. This should, however, not be regarded as a
limitation of SIMF: The method is designed to be applied foremost to {\em
fractal} structures, and these are known to have ``arms'' that rarely exceed
4-5 atoms across \cite{Behm,Kern:2}. If specular scattering from the interior
of ad-islands is important, the adatom system is likely to contain large
compact regions and should probably be viewed as a ``fat fractal''
\cite{Umberger}, i.e., a fractal with non-zero Lesbegue measure. Such fractals
are characterized by their ``exterior dimension'' \cite{Grebogi}, i.e., the
fractal dimension of the rim of the islands. But since it is the rim which
accounts for the off-specular intensity, one can still apply SIMF, and find
that it measures to a good approximation just this exterior dimension. Thus
the method is seen to naturally provide the proper characterization of fat
fractals as well. A point of concern regarding Eq.(\ref{eq:I-system}) is that
it implicitly employs the ``geometric overlap principle'' \cite{Comsa:book} of
adatom CSs. This has been criticized for adatom {\em pairs}
\cite{Peppino}. For other systems \cite{Comsa:book,Benny:3,comment1} it is
strongly corroborated by test cases.  We accepted it here, by assumption, as a
method valid for all systems, for sufficiently low coverages. For some
systems, at high coverage, systematic higher-order corrections to
Eq.(\ref{eq:I-system}) may be necessary, which in our belief can be
introduced. However, in this study we shall not deal with these complications.

\section{Results and Discussion}
We proceed to present the results of our analysis of combined experimental and
simulation data. To simulate (presently unavailable) experimental data, we
performed He scattering calculations within the Sudden approximation (SA)
\cite{Benny:review2} on configurations of Ag/Pt(111) generated by kinetic
Monte Carlo (KMC) simulations \cite{me:Heptamers}, in accordance with the
Deposition, Diffusion and Aggregation (DDA) scheme \cite{Stanley}. This scheme
has been very successful in modeling thin-film-deposition experiments. A
typical configuration generated by our simulations at a relatively high
coverage of 27\% is
shown in the inset of Fig.\ref{fig:D}, and is reminiscent of the STM
topographs of Refs.\cite{Behm,Kern:2}. Experimental CS data was available for
scattering from a single Ag/Pt(111) adatom \cite{Zeppenfeld}. The He/Ag/Pt
potential (to be published separately in Ref.\cite{me:Ag-systems}),
unavailable at present from other sources, was obtained by fitting SA-CS
calculations to this data. (To test its reliability, the same potential was
used independently to reproduce an experimental angular intensity distribution
for scattering from a system of polydispersed, small Ag/Pt(111) clusters,
which it did with impressive accuracy \cite{me:Ag-systems}). The fit is shown
in Fig.\ref{fig:single-cs}. Clearly, this potential is approximate only.
However, we believe it is realistic enough for the purpose of simulating
scattering data. In any case, the method does not depend on any specific
potential: The ultimate test of SIMF will have to be experimental. The
single-adatom CS, $\Sigma_s(k_z)$, shown in Fig.\ref{fig:single-cs}, was
extracted from the specular intensity according to Eq.(\ref{eq:I-single}),
yielding $R_s(k_z)$. As seen, beyond $k_z^m \!\approx\! 10$bohr$^{-1}
\!\approx\! 10^3K$, $\Sigma_s(k_z)$ tends to a minimal limiting value
$\Sigma_s^m \!=\! \pi R_m^2$, with $R_m \!=\! 4.55$\AA. The Pt(111)
lattice-constant is $2.77$\AA, thus the CSs of nearest-neighbor Ag adatoms
always overlap, and the set appears as a continuum to the incoming He. This is
shown in the inset of Fig.\ref{fig:D}, where a circle of radius $R_m$ is drawn
around each nucleus, and the resulting contour is shown. This is an
approximate view of the contour of constant electron density observed by  He
atom incident at $k_z > k_z^m$, and of the fractal whose dimension is measured
by SIMF. The
adatom system CS, $A_s(k_z)$, was found according to Eq.(\ref{eq:I-system})
from the specular intensities calculated at various coverages, shown in
Fig.\ref{fig:spec-islands}.

Once the functions $R_s(k_z)$ and $A_s(k_z)$ are known, the FD can be found.
The algorithm we propose is based on the Minkowski \cite{Mandelbrot}
definition. The Minkowski cover $M(R)$ of a set $X$ is obtained by placing
circles of radius $R$ centered at every point in $X$. One then calculates the
area $A(R)$ of the union of all circles. The FD is found in practice as the
slope at small $R$ values of the log-log plot

\begin{equation}
\left\{ \log {1 \over R}, \log {{A(R)} \over R^2} \right\},
\label{eq:log-log}
\end{equation}

\noindent which presumes that a constant slope (i.e., scaling) is observed over
at least one decade \cite{Pfeifer-in-Avnir:book}. In the present context, a
natural choice of the set $X$ is the union of adatom CSs at the high
$k_z$/minimal radius ($R_m$) limit, shown in the inset of Fig.\ref{fig:D}. Since in this limit the CS is determined
by the repulsive part of the potential, semi-classically, the CS circumference
corresponds, through Eq.(\ref{eq:Norskov}), to a contour of constant electron
density. Therefore, this choice yields the FD corresponding to such contours
of the Ag system, where a high-energy He atom would be stopped. This is an
inherent property of the adatom system, uniquely accessible to He. Other
probes (e.g., STM) would measure the FD of another feature of the adatom
system. The remarkable property of He scattering, is that it is capable of
providing the Minkowski cover of the adatom set, through a variation of the
incidence energy, which is equivalent to a change of the scattering-radius
$R_s$. Thus, the process summarized by Eq.(\ref{eq:log-log}) can be carried
out in an He scattering experiment in two steps: First, the CS of an
individual Ag adatom is measured over a wide range of incidence wavenumbers,
at least as far as the saturation point $k_z^m$. This then yields $R_s(k_z)$,
with the minimal radius measured as $R_m = R_s(k_z^m)$. Second, $A_s(k_z)$ is
measured under the same conditions of variable incidence
wavenumbers. Eliminating $k_z$ between $A_s$ and $R_s\!=\!  R_m\!+\!R$ yields
the scattering-determined area of the Minkowski-cover, $A_s(R_s)$. From here
the calculation of the FD proceeds according to Eq.(\ref{eq:log-log}). (Note
that it is the ``Minkowski-radius'', $R$, and not the full cross-sectional
radius $R_s$, which appears in Eq.(\ref{eq:log-log})). We applied this
procedure, for step 1, to the CS fitted to the experimental Ag/Pt(111) data
(Fig.\ref{fig:single-cs}), and for step 2, to our scattering simulation data
at various coverages. The results are shown in Fig.\ref{fig:log-log}. It is
noteworthy that 2 decades of linear, scaling behavior are observed.  {\em The
existence of a large scaling regime is an ``empirical'' proof of the
fractality of our systems, and the ability of He scattering to measure it}. It
is tempting to compare the scattering results to a simple geometrical model,
where the Minkowski procedure is applied to circles of radius $R_m$, centered
at the adatom positions. However, such a naive approach is misleading: The
quantum CS has no simple geometrical counterpart. The CS is ``fuzzy'', i.e., a
He atom is scattered off-specularly from within the CS only with a
probability. Work incorporating a distribution of geometrical radii, allowing
a comparison with the CS, is currently in progress.

The corresponding slopes, or FDs, are displayed as a function of coverage in
Fig.\ref{fig:D}. The FD rises monotonically with coverage and is close to the
embedding-space dimension of 2. This high FD is expected, if the large value
of the individual CS (Fig.\ref{fig:single-cs}) is taken into account: As seen in the inset of Fig.\ref{fig:D}, due to this large CS the contour
has a rather smooth structure (compared, e.g., to a typical, highly ramified,
DLA cluster), thus enclosing an almost ``ordinary'' 2D set. The
non-universality of the FD (i.e., its dependence on coverage) is a definite
prediction of the theory, and differs significantly from the well-known result
for {\em individual} DLA clusters, the only fractal systems observable so far
by STM \cite{Behm}, whose FD$\approx\!$ 1.7 \cite{Witten} is
universal. However, it does agree with the results obtained by Jensen et
al. \cite{Stanley} for the DDA model: There too, a strong dependence of the FD
on coverage is observed. More importantly, as previously mentioned, $R_m > a$,
the Pt(111) lattice constant. This implies that the geometry probed by the He
beam is much less porous that that of DLA clusters. Gaps in the latter can
easily be smoothed out by overlapping CSs, and disjoint DLA clusters may in
fact become one from the He point of view. The rise in FD can be understood as
an increasingly larger portion of the surface being covered by overlapping
CSs, as the coverage is increased. Taken together, these observations should
explain why the geometry of constant electron density contours, probed by a He
beam, has a very different fractal nature from that of DLA clusters. We expect
actual He scattering experiments to produce similar result to
Figs.\ref{fig:single-cs}-\ref{fig:D}.

\section{Concluding Remarks}
In summary, a simple new procedure for the measurement of FD of disordered
submonolayers by He scattering has been proposed. First the specular intensity
as a function of incidence wavenumber $k_z$ for a single adsorbate must be
measured, from which the scattering-radius $R_s$ is determined according to
Eq.(\ref{eq:I-single}). Then the specular intensity is measured over the same
incidence wavenumber range for the adatom system of interest, and its CS,
$A_s$, is found according to Eq.(\ref{eq:I-system}). Eliminating the
wavenumber between these two quantities, the function $A_s(R_s)$ is plotted on
a log-log scale according to Eq.(\ref{eq:log-log}). The slope of this plot is
the FD. If the adatom system contains large compact regions, the calculated FD
should be interpreted as the exterior dimension of a fat fractal.

SIMF measures the FD of constant electron density contours of an adsorbed
metal system. This observation raises some interesting questions for future
research, which is beyond the He scattering context: For example, what is the
effect of this fractality on bulk properties, such as surface conductivity and
elasticity?

Finally, the present article does not deal with the effect of adatoms adsorbed
near step edges. These adatoms have different CSs from those adsorbed on
terraces, and hence require a separate calibration of $\Sigma_s(k_z)$. This
issue is currently under investigation. An initial experimental effort to
implement SIMF for Ni/Cu(111) is under way \cite{Rosenfeld:private}.

\acknowledgements
This research was supported by the German-Israeli Foundation for Scientific
Research (G.I.F.), under grant number I-215-006.5/91 (to R.B.G.). Part of this
work was carried out with support from the Institute of Surface and Interface
Science (ISIS) at UC Irvine. The Fritz Haber Center at the Hebrew University
is supported by the Minerva Gesellschaft f\"{u}r die Forschung, Munich,
Germany. We are grateful to Prof. G. Comsa and Dr. P. Zeppenfeld for
stimulating discussions, and thank Profs. D. Avnir, O. Biham and G. Vidali,
and D. Thimor for helpful comments.

\begin{figure}
\caption{Cross section of single Ag atom on Pt(111) as a function of incidence
wavenumber. Circles are experimental results \protect\cite{Zeppenfeld},
accurate to within 20\%. Dots connected by solid line are calculated CS
values, using an LJ 6-12 potential with parameters fitted to the experimental
data. The CS saturates at $k_z^m \protect\approx 10$ bohr$^{-1}$. Small
interference oscillations due to particles striking the flat surface and the
adatom tops can be observed. These measurements and simulations constitute
step 1 in the SIMF algorithm.}
\label{fig:single-cs}
\end{figure}

\begin{figure}
\caption{Specular intensity of He scattered by fractal adatom islands of
Ag/Pt(111), for coverages of 3\protect\%-27\protect\%, averaged over 30
configurations at each coverage. Interference oscillations increase in
magnitude with the coverage. The specular intensity overall increases with
wavenumber, reflecting the decrease in total cross section of the Ag islands.
These simulations constitute step 2 in the SIMF algorithm.}
\label{fig:spec-islands}
\end{figure}

\begin{figure}
\caption{Plots of $\log[A_s(R_s)]/R^2$ vs $\log(R_m/R)$, for various coverages,
used to determine the FD (note that the oscillations from
Figs.\protect\ref{fig:single-cs},\protect\ref{fig:spec-islands} have cancelled
out). Two decades of scaling behavior are observed, indicating that He is
indeed capable of measuring the FD of the scattering object.}
\label{fig:log-log}
\end{figure}

\begin{figure}
\caption{He scattering prediction of fractal dimension as a function of
coverage, obtained from the log-log plots in Fig.2. Error bars are due
to linear regression. Inset: Dots are nuclei positions in typical
configuration obtained by KMC simulations on a Pt(111) hexagonal  
surface (periodic boundary conditions apply). Coverage is 27\%. Contour
represents view of electron density by He atom incident with $k_z>k_z^m$. That
is, each adatom is drawn to scale with a cross-sectional radius of
$R_m=4.55\AA$. According to SIMF, this fractal has a dimension of 1.98.}
\label{fig:D}
\end{figure}

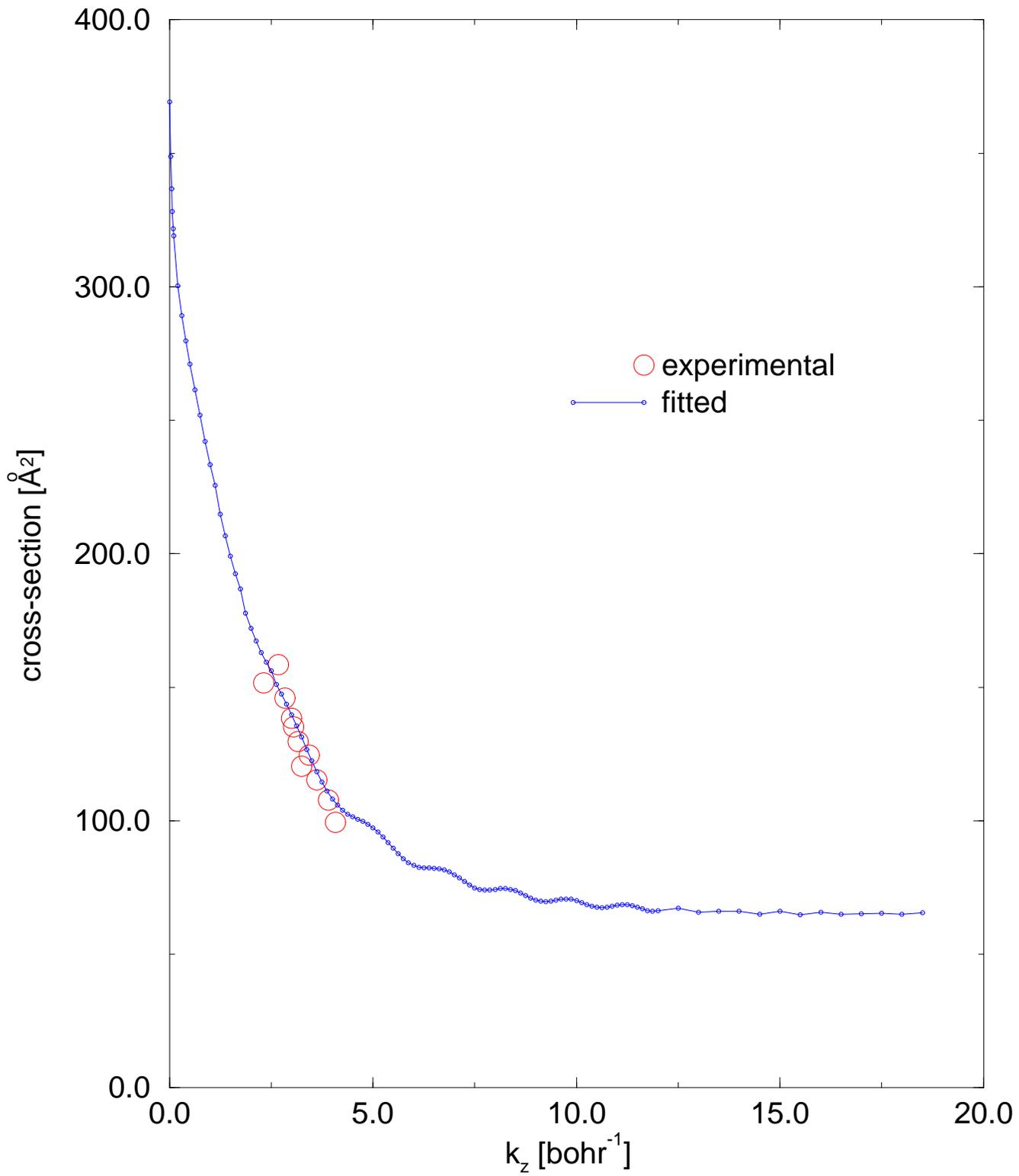
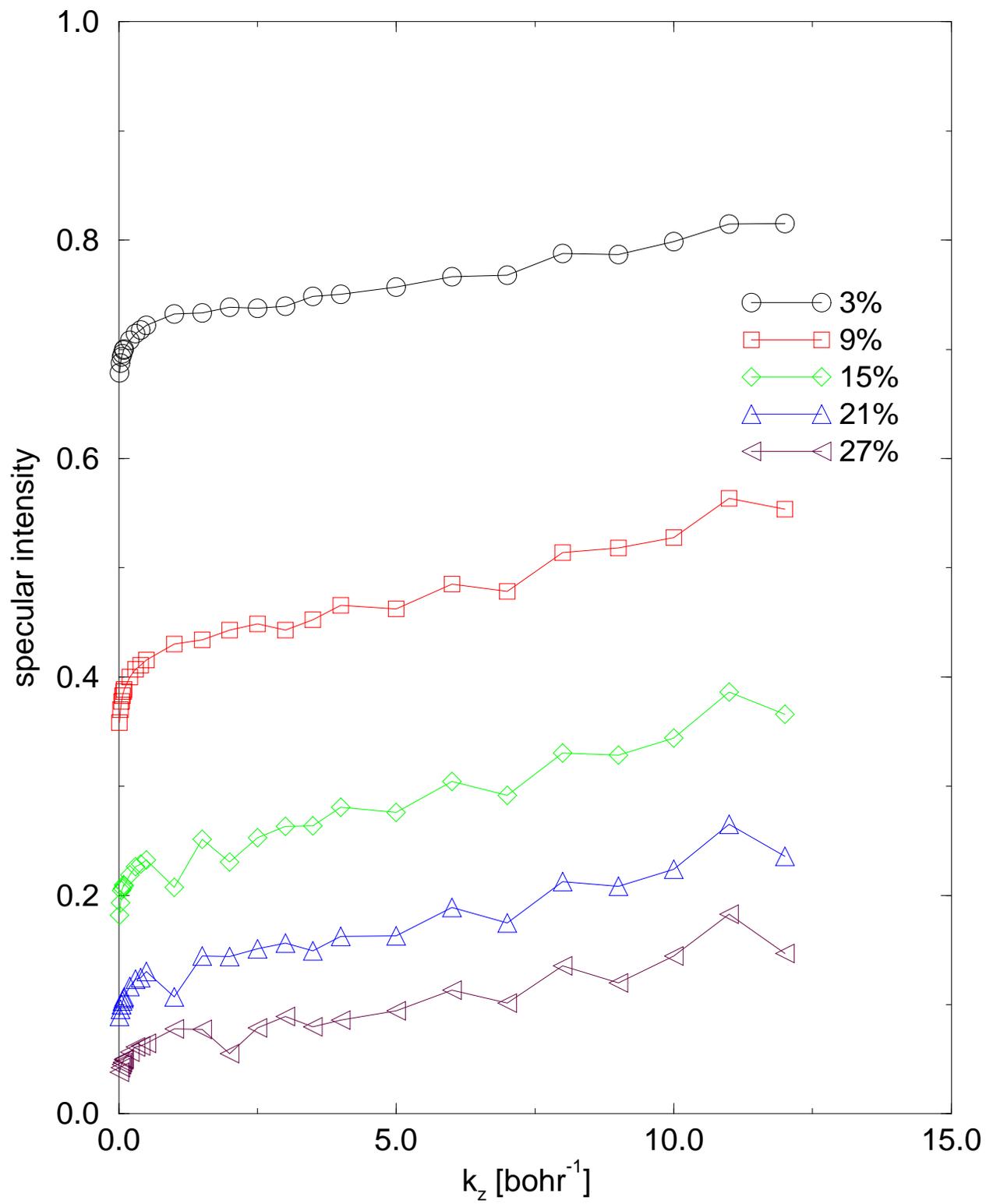
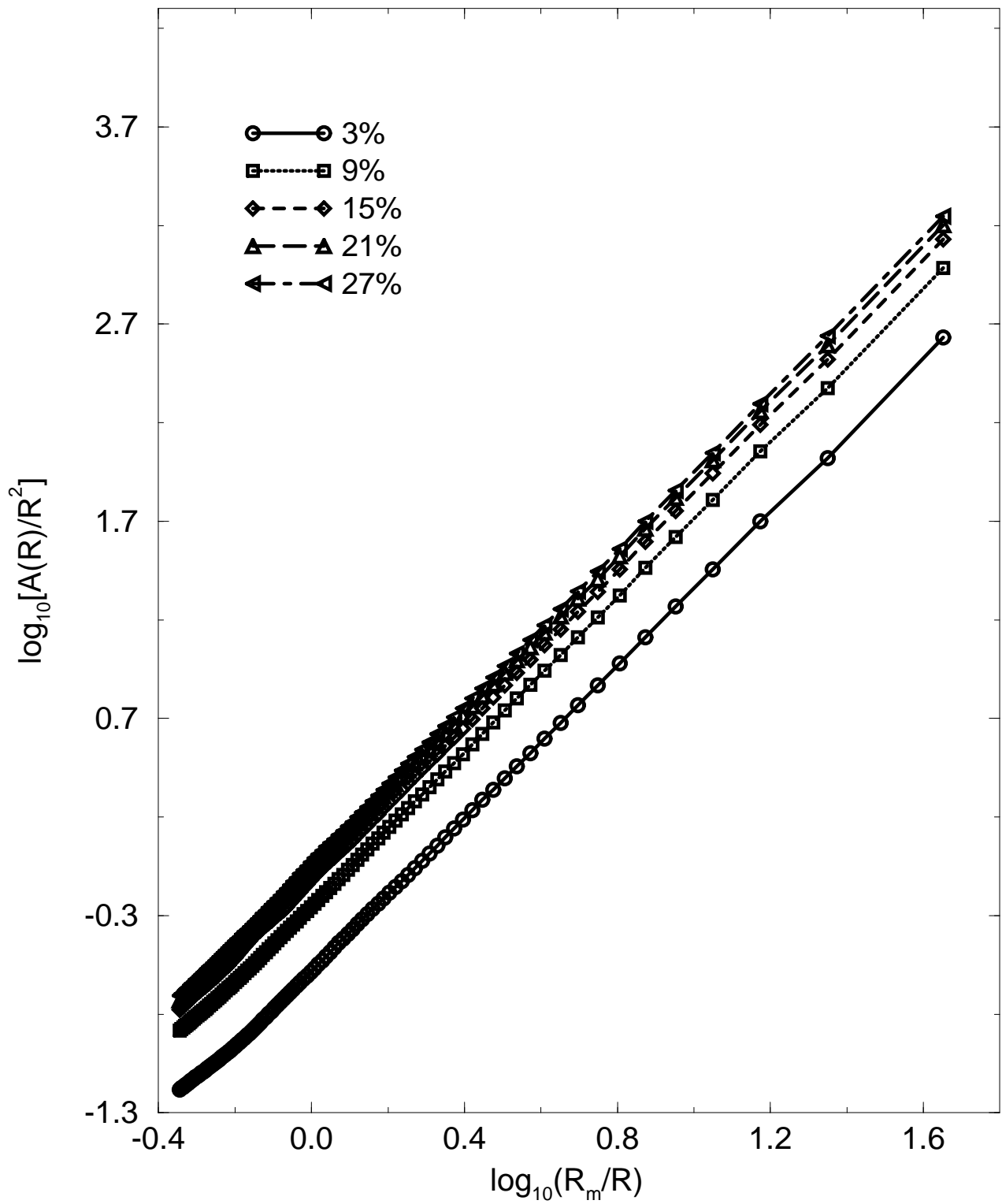
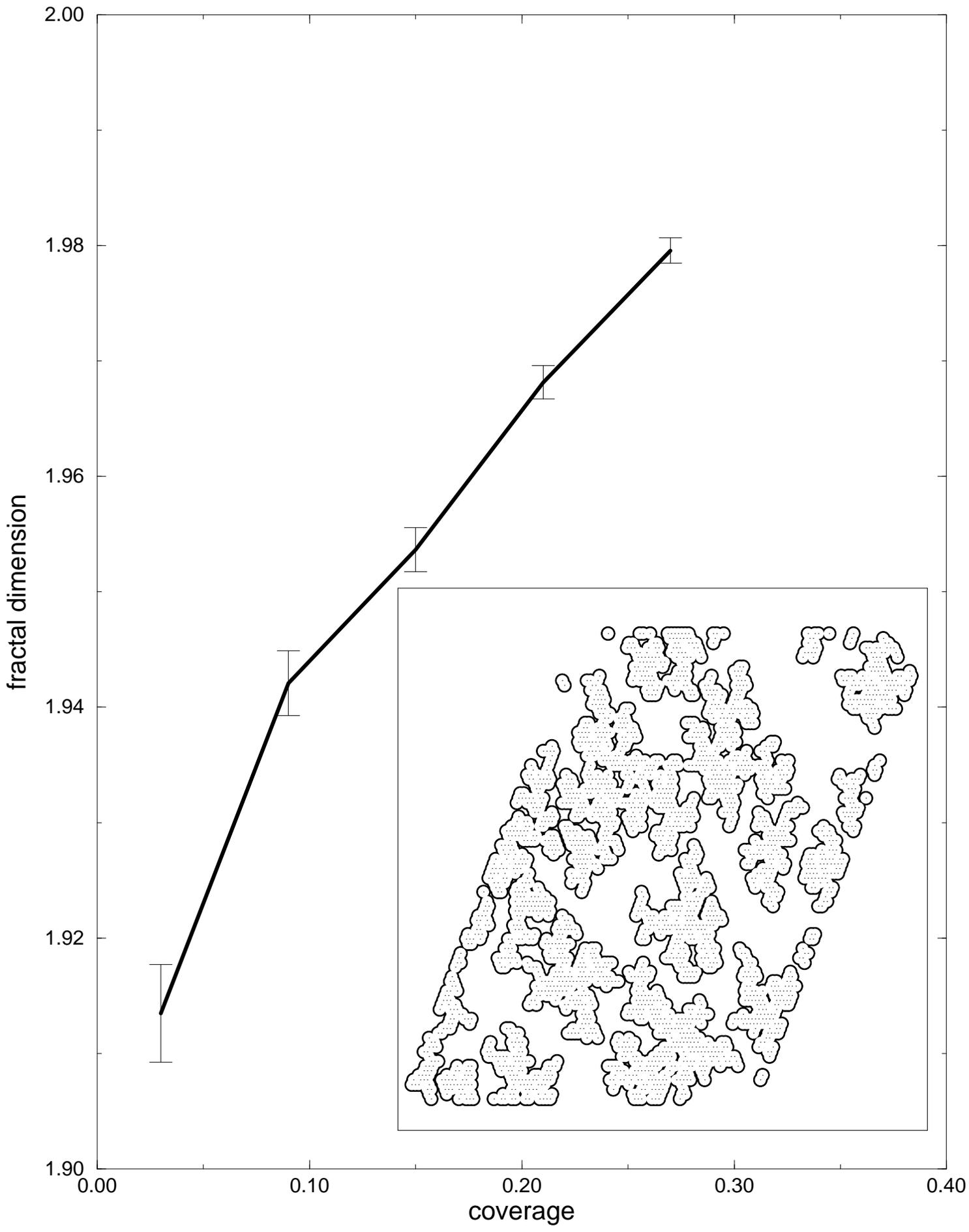
\end{document}